\documentclass[11pt]{article}

\usepackage[margin=1in]{geometry}
\usepackage[utf8]{inputenc}
\usepackage[T1]{fontenc}
\usepackage{hyperref}
\usepackage{url}
\usepackage{booktabs}
\usepackage{amsfonts}
\usepackage{amsmath}
\usepackage{nicefrac}
\usepackage{microtype}
\usepackage{xcolor}
\usepackage{graphicx}
\usepackage{enumitem}

\title{Infrastructure-Aware Runtime Protection of\\
Co-Deployed AI Services at the Industrial Edge}

\date{}

\author{%
  Nitya Sathyavageeswaran \quad Dibyajyoti Nath \quad Ziran Min \quad Pavan Athreya Narasimha Murthy
}

\begin{document}
\maketitle

\begin{abstract}

In modern production lines, a single edge device simultaneously runs multiple AI services like a torque-control model, a vision inspector, a predictive maintenance routine. All these services run on the same CPU and same shared network segment. When the shared network segment saturates, the services do not crash. The services just produce outputs that are inaccurate or stale, and no longer reflect the plant's conditions. Our paper targets this class of silent, partial failure scenarios in Industrial Edge Systems. 

We treat each co-deployed service as an autonomous agent with its own inputs, outputs, and survival urgency.
We introduce the \textbf{Runtime Criticality Score (RCS)}, which each service agent uses to express at any moment, its intrinsic utility and survival urgency under resource constraints.  
The score is computed from six normalized
dimensions covering physical consequence, inference quality, input freshness, recent
availability stress, actuator coupling, and available protective actions on the host. Its dimension weights are adjusted online through a lightweight
Bayesian update driven by the type of disturbance currently observed, so the same service is ranked differently under a network delay than under thermal throttling. Agents publish their RCS to peers over an open, peer-to-peer discovery
channel, and use it to negotiate three things without a central controller: (i) local
threshold enforcement on their own host, (ii) migration to a willing peer when the current host
can no longer protect them, (iii) reassignment of the coordination role when the incumbent
coordinator is itself impaired. Beyond the immediate benefit of catching silent degradation on the factory floor, the
framework contributes a concrete instance of decentralized, physically grounded agent
coordination. Agents belonging to different owners can advertise capabilities, bid for
protection, and reorganize across an open network.
\end{abstract}

\section{Introduction}

 Most AI services deployed at the industrial edge are designed for isolation~\cite{shi2016edge}. Each service is built, tested, and monitored as if it were the only workload on the device. In practice, the device is shared. A single edge box on a production line will often carry a control model, one or two vision inspectors, a predictive maintenance routine, and a telemetry collector, all sharing the same CPU, memory, and network link. Under contention these services rarely stop. A maintenance model keeps issuing scores even after competing traffic has delayed its sensor updates, and a quality inspector keeps labeling parts after its processor has begun to throttle. From the outside the system still looks operational. The failure only becomes visible once a defect slips through, a line halts, or a control decision is made on an inference that was no longer trustworthy. Current monitoring does not catch this kind of silent, partial failure, because each service is observed on its own and none of them looks broken in isolation. Existing orchestration and watchdog systems on the edge track liveness, CPU, and memory per container~\cite{xiong2018extend,hong2019resource}. We treat each co-deployed service as an autonomous agent with its own inputs, outputs, and survival urgency. They do not weigh one degrading service against another. What is missing is a way for these agents to rank themselves against one another while they are all degrading. 
 Existing tools produce no score or signal that reflects the physical consequences of acting on a bad output, the state of the inference itself, among other factors~\cite{carvalho2019pm, fox1999harvest}. Without such a signal, no agent, and no autonomous supervisor, knows which service to protect. 
 
 We introduce the \textbf{Runtime Criticality Score} (RCS), a per-service priority evaluated on each device and shared across a thin distributed runtime layer. Dimension weights are updated online through a lightweight Bayesian rule whose posterior is conditioned on a disturbance type. The posterior is conditioned on the disturbance type currently observed, specifically network congestion, thermal throttling, or memory pressure, inferred from network latency, CPU temperature, and local queue depth. The disturbance taxonomy is configurable, and operators may define additional types to match their deployment. Priority-ranked migration and coordinator fail-over then let the most consequential agents be protected first, even when the coordinator arranging that protection is itself impaired. 
 
 The paper makes four contributions: \begin{itemize}[leftmargin=*, nosep]
 \item \textbf{A unified priority signal.} The RCS combines six normalized dimensions that jointly capture physical consequence, inference quality, input freshness, recent availability stress, actuator coupling, and residual protective headroom on the host. 
 \item \textbf{Condition-driven online adaptation.} Dimension weights are updated by a lightweight Bayesian rule keyed to the currently inferred disturbance type, so the ranking adapts to infrastructure change without retraining the services.
 \item \textbf{Priority-ranked coordination across peers.} A distributed runtime layer uses the RCS to drive local threshold enforcement, migration of workloads in decreasing score order, and consequence-weighted fail-over of the coordination role when the primary host is itself impaired. 
 \item \textbf{An open, multi-owner deployment model.} Agents publish their RCS over a peer-to-peer discovery channel and coordinate without a shared trust root. The present system trusts each agent to report its score honestly. 
 \end{itemize}

\section{Related Work}

Prior work on edge deployment, inference adaptation, service migration, and agent coordination each addresses part of the problem we target. To our knowledge, no existing work brings them together under a single runtime signal.

Fox and Brewer studied trading data completeness for availability under stress~\cite{fox1999harvest}. They show that a system should shed lower-priority work to preserve core function. Their framework, however, is defined over a single application's aggregate output and provides no mechanism to rank several simultaneously degraded services against one another.

Latency-sensitive IoT processing at the network edge has been extensively studied~\cite{shi2016edge, mahmud2017fog, mach2017mobile}. These works take the presence of multiple co-deployed services for granted but treat contention between those services as a background condition rather than the primary object of study.

Adapting DNN inference pipelines to fluctuating network conditions and monitoring industrial equipment through machine learning have each been studied in depth~\cite{li2019edgeai, carvalho2019pm}.  Li et al.'s Edgent~\cite{li2019edgeai} adapts a single on-device DNN pipeline to network fluctuation locally but never arbitrates between it and a co-located maintenance model competing for the same resources. Carvalho et al.~\cite{carvalho2019pm} survey the breadth of machine learning methods deployed for predictive maintenance on the shop floor, illustrating how operationally consequential a single degraded pipeline can be, but neither work addresses simultaneous degradation across services.

Service migration in mobile edge-clouds has been formalized as a Markov decision process~\cite{wang2015migration}. Migration is initiated by user mobility rather than service criticality, and the cost model does not distinguish a bad recommendation from a wrong actuator command, so the policy cannot be reused when the question is which service to move to protect physical safety.

Autonomous agent coordination, task reallocation, and leader re-election in decentralized collectives have been studied across a range of settings~\cite{wooldridge1995intelligent, dorri2018multi, stone2000multiagent, korsah2013comprehensive, ongaro2014search, demers1987epidemic}. We treat each co-deployed service as an agent in this sense directly. None of this literature, however, addresses priority ordering when multiple agents degrade simultaneously under shared infrastructure stress. The weight-adaptation mechanism connects to multi-agent reinforcement learning~\cite{busoniu2008comprehensive, sutton1998reinforcement}, which we identify as a direction for future work. The agent registration schema follows the structure of the Model Context Protocol~\cite{anthropic2024mcp}, making capability declarations checkable rather than self-reported.

All these papers look at a single service or a single device. None asks what happens when an output produced under degradation is acted upon in the plant. The RCS framework we introduce  produces a cross-service, condition-adapted ranking shared across peers, and uses it to drive local protection, cross-device migration, and coordinator fail-over under one signal. This matters most where equipment from several vendors coexists on the same floor and no single administrative domain can be assumed to remain healthy on behalf of the rest.

\section{The RCS Framework} 

Each co-deployed service is an agent: it has its own inputs, outputs, and a stake in remaining operational. The framework is built around one question. Among all agents running on a contested device, which one has the most to lose if it is not defended in the next control cycle? The pieces below, registration, scoring, local enforcement, and peer coordination, are the machinery that answers that question and acts on it. 

\subsection{System Architecture} 

\begin{figure}
    \centering \includegraphics[width=0.8\linewidth, height=0.28\textheight, keepaspectratio]{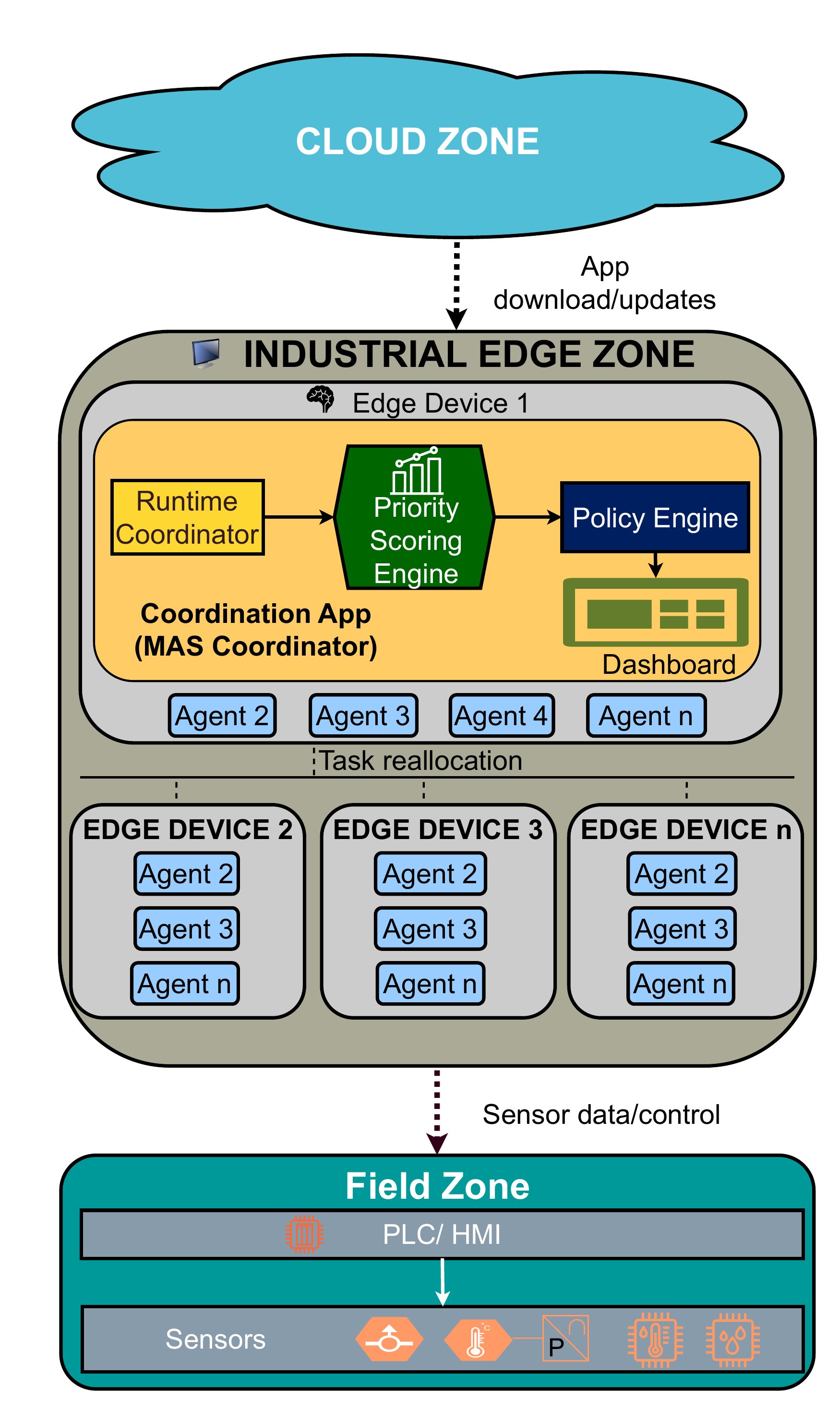} \caption{Three-zone architecture: the Cloud Zone distributes updates only; the Industrial Edge Zone runs agent coordination, RCS computation, and priority-ordered migration across peer devices; the Field Zone contains the Programmable Logic Controllers (PLCs), Human-Machine Interfaces (HMIs), and sensors that ground the RCS in physical consequence.}\label{fig:arch} 
\end{figure}

The deployment is organized into three zones as shown in Figure~\ref{fig:arch}. The \emph{Cloud Zone} distributes model updates and configuration but plays no part in runtime decisions. The \emph{Industrial Edge Zone} is where all runtime coordination happens. Each device runs one or more agents and a thin coordination sub-layer that computes RCS locally, exchanges it with peers, and acts on threshold crossings. One device holds the coordinator role at any given time, but every peer keeps enough state to assume that role on fail-over. The \emph{Field Zone} contains the sensors and actuators the agents actually control, and it is the reason the RCS is grounded in physical consequence rather than software liveness alone. 

\subsection{Agent Registration and Runtime Handling Classes} 

An agent becomes visible to the coordination layer only after it registers. At registration, each agent declares four things the infrastructure cannot infer: its workload type, the physical consequence severity of a degraded output, the maximum input age that still yields trustworthy inference, and the protective actions its host can perform on its behalf. From these declarations the coordinator assigns the agent to a \emph{runtime handling class}. \emph{Safety-critical} agents drive actuators directly and their degradation can cause immediate harm. \emph{Supervisory} agents shape process outcomes one step removed from the actuator. \emph{Analytical} agents produce advisory outputs and are the natural candidates for graceful degradation under resource pressure. The handling class determines both the aggressiveness of the response to a threshold crossing and the baseline weights used in the RCS computation, so an agent controlling a high-pressure valve is never treated the same way as one that recommends a set-point adjustment.

\subsection{Runtime Criticality Score} For each registered agent $s$, the RCS is a weighted sum of six normalized dimensions evaluated at every monitoring tick: \begin{align}
    \mathrm{RCS}_s(t) = \sum_{i=1}^{6} w_i(t)\, d_{i,s}(t) \label{eq:rcs} .
\end{align} 
where  $d_{i,s}(t) \in [0,1]$ is the normalized value of dimension $i$ for agent $s$ at time $t$, and $w_i(t)$ is the corresponding dimension weight, updated online based on the disturbance type currently observed on the host.
The six dimensions are: physical consequence severity (\textbf{PCS}), inference quality drift from the agent's baseline distribution (\textbf{IFS}), input staleness relative to the freshness bound declared at registration (\textbf{IFR}), recent history of contention on the host (\textbf{HAS}), actuator coupling depth so that a one-cycle path to a valve scores above a many-hop pipeline (\textbf{ACD}), and residual device control depth as the host exhausts its isolation and throttling options (\textbf{DCD}). Each dimension is normalized to $[0,1]$ at registration so that quantities with different units become directly comparable across agents.\footnote{The three handling classes align with IEC~62443 zoning and IEC~61511 SIL categories.} 
These six dimensions are chosen to span the full causal chain from sensor input to physical output: consequence severity, model state, input freshness, host stress history, actuator coupling, and residual protective headroom. Together they capture every point at which silent degradation can originate or amplify before reaching the plant floor.
\paragraph{Condition-aware weight adaptation.}We model three disturbance types that together cover the primary failure modes observed in shared edge deployments: network congestion, thermal throttling, and memory pressure. Let $c \in \{\text{network},\, \text{thermal},\, \text{memory}\}$ denote the active disturbance type. This taxonomy is configurable; operators may define additional types to match their deployment.

Let $o_t = (l_t, \tau_t, q_t)$ be the observation vector, where $l_t$ is network latency, $\tau_t$ is CPU temperature, and $q_t$ is the number of unprocessed messages pending on the device. Each device's coordination layer maintains a Bayesian posterior over disturbance type $c$ given $o_t$ which is given by
\begin{align}
    p(c \mid o_t) \propto p(o_t \mid c)\, p(c). \label{eq:posterior}
\end{align}

\noindent The active weight vector $w(t)$, is the posterior-weighted mixture of class-specific templates $w^{(c)}$, fixed at deployment and estimated once from a short calibration run and is given by
\begin{align}
    w(t) = \sum_{c} p(c \mid o_t)\, w^{(c)} .\label{eq:weights}
\end{align}

\noindent Under network delay, $w_{\mathrm{IFR}}$ rises as stale inputs become the dominant threat. Under thermal throttling, $w_{\mathrm{IFS}}$ and $w_{\mathrm{DCD}}$ rise instead. 

\subsection{Threshold-Based Local Enforcement} Thresholds are parameterized per handling class. A low-severity crossing triggers local resource protection such as CPU reservation. A mid-severity crossing notifies the operator with the current RCS and the dimensions contributing most to it. A high-severity crossing initiates a safe-state transition: lower-priority agents may be suspended, the affected agent switched to a degraded-but-safe mode, or migrated to a peer. Every response is keyed to the same score and class, so behavior is proportionate and traceable. 

\subsection{Distributed Coordination: Migration and Fail-over} Each device maintains a \emph{membership record} for reachable peers, tracking operational health, spare capacity, and coordinator eligibility. 
\paragraph{Priority-ordered migration.} 
Migration is triggered when the current host can no longer protect an agent at its required RCS level. Among agents crossing the migration threshold, the highest scorer moves first. For each candidate peer $k$, a projected post-migration score is computed holding PCS, IFS, and IFR fixed and recomputing HAS, ACD, and DCD from the target's resource state:
\begin{align} 
    \widehat{\mathrm{RCS}}_s^{(k)}(t) = \sum_{i=1}^{6} w_i(t)\, \hat{d}_{i,s}^{(k)}(t) \label{eq:proj} 
\end{align} 
where $\hat{d}_{i,s}^{(p)}(t)$ is the projected value of dimension $i$ for agent $s$ on peer $k$. The peer maximizing $\widehat{\mathrm{RCS}}_s^{(k)}$ is selected, with ties broken by largest residual DCD headroom. If no peer exceeds the migration threshold, the high-severity local response takes over.
\paragraph{Consequence-weighted coordinator fail-over.} If the primary coordinator becomes degraded or unreachable, an eligible peer takes over. The successor inherits the membership record and current RCS values, so coordination continues rather than restarts. This removes the coordinator as a single point of failure and removes the assumption that any single administrative domain must remain healthy on behalf of the rest.

\section{Discussion and Conclusion}

The RCS tells a contested edge device which agent to defend first, adapts 
that decision to the infrastructure disturbance in progress, and lets 
peers reorganize when the local host or coordinator can no longer carry 
the load. Each service-as-agent expresses survival urgency in a form 
peers can act on, without a shared owner, trust root, or central 
controller.

Two limitations remain open. The disturbance taxonomy and weight templates are fixed at deployment; online updates would extend adaptation without changing the interface. Also, the protocol trusts each peer to report its RCS honestly; defending against strategic reporting and adversarial peers is hard. A full hardware evaluation under injected network, thermal, and memory pressure remains an important next step, and represents a physically consequential testbed for coordination and trust mechanisms that must eventually operate across administrative domains.

\bibliographystyle{unsrt}
\bibliography{references}
\end{document}